\documentclass[3p,times,twocolumn]{elsarticle}
 \biboptions{comma,sort&compress}
 
\usepackage{graphicx}
\usepackage{here}
\usepackage{ecrc}
\usepackage{url}

\volume{00}

\firstpage{1}

\journalname{Nuclear and Particle Physics Proceedings}

\runauth{}

\jid{nppp}

\jnltitlelogo{Nuclear and Particle Physics Proceedings}

\usepackage{amssymb}

\usepackage[figuresright]{rotating}

\begin{document}

\begin{frontmatter}


\title{Tetra-quarks, penta-quarks and the like: old and new views
 $^*$}
\cortext[cor0]{Talk given at 23$^{rd}$ High Energy Physics International Conference in Quantum Chromodynamics, QCD20 (35$^{th}$ anniversary), October 27-30, 2020, Montpellier - France.\\
\indent Report Number CERN-TH-2020-197}

 \author[label1,label2]{G.\ C.\ Rossi\corref{cor1}}
\ead{rossig@roma2.infn.it}
\address[label1]{Universit\`a di Roma {\emph {Tor Vergata}} and INFN - Sezione di Roma {\emph {Tor Vergata}}, Via della Ricerca Scientifica, 00133 Roma, Italia}
\address[label2]{Centro Fermi - Museo Storico della Fisica e Centro Studi e Ricerche ``E.\ Fermi'', Piazza del Viminale 1, 00184 Roma, Italy}
\cortext[cor1]{Speaker}

\author[label3,label4]{G.\ Veneziano}
\ead{gabriele.veneziano@cern.ch}
\address[label3]{Coll\`ege de France, 11 place M.\ Berthelot, 75231 Paris, France}
\address[label4]{Theory Department, CERN, CH-1211 Geneva 23, Switzerland}

\pagestyle{myheadings}
\markright{ }
\begin{abstract}
In this talk, after a short overview of the history of the discovery of tetra-quarks and penta-quarks, we will discuss a possible interpretation of such states in the framework of a 40-years-old ``string junction'' picture that allows a unified QCD description of ordinary mesons and baryons as well as multi-quark resonances.
\end{abstract}


\end{frontmatter}

\section{A historical introduction}

The S(1930) bump identified in the early seventies as a very narrow structure ($\Gamma\!\sim \!4$~MeV) in $\overline p p$ and $\overline p d$ total cross sections~\cite{CAR} was the first resonance tentatively interpreted as a tetra-quark~\cite{Rossi:1977cy,Rossi:1977dp}. Despite a number of experiments confirming it~\cite{Chaloupka:1976yj,Bruckner:1976pu,Sakamoto:1979je}, S(1930) finally disappeared from the radars after the unsuccessful high statistics search carried out in ref.~\cite{AlstonGarnjost:1975vf} and the systematic investigation performed at LEAR~\cite{LEAR}.

Soon after in 1980 in the review paper~\cite{Montanet:1980te} penta-quarks, i.e.\ states with a $[qq\bar q qq]$ structure, were conjectured to exist on the same footing as tetra-quarks. However, it wasn't until 2003 that the discovery of a narrow ($\Gamma\!\!\sim\! \!10$~MeV) penta-quark candidate $\Theta_5(1540)$ with $[d d\bar s u u]$ flavour content was claimed by the LEPS Collaboration~\cite{Nakano:2003qx} (see also the list of references from [1] to [12] collected in~\cite{Csikor:2005xb}). Unluckily, this state was not later found in inclusive experiments (see~\cite{Bai:2004gk} and the references from [14] to [20] in~\cite{Csikor:2005xb}). A similar fate had the exotic conjectured penta-quark partner $\Xi^{--}(1860)$ with $[ds\bar u ds]$ flavour content reported by the NA49 collaboration~\cite{Alt:2003vb} and the charmed state with $M\!\sim \!3100$~MeV, $\Gamma\!\sim \!12$~MeV and $[uu\bar c dd]$ flavour content seen in~\cite{Aktas:2004qf}.
In the meantime the light scalar mesons ($f^\circ,\sigma^\circ, a,\kappa$) mass puzzle was solved~\cite{Maiani:2004uc} by picturing these states as diquark--anti-diquark~\cite{Jaffe:1976ih} composites.

Starting from 2013 a plethora of heavy, narrow mesonic states, termed $X,Y,Z$ (among which the extremely narrow exotic $X(3872)$ resonance~\cite{Choi:2003ue,Aaij:2020qga}~\cite{Maiani:2004vq,Matheus:2006xi}) and interpreted as tetra-quarks, were identified by Belle and LHCb~\footnote{Useful, though not so recent, compilations appeared  in~\cite{Eidelman:2012vu}~\cite{Bodwin:2013nua}.} and in many cases confirmed by more than one experiment. The discovery of narrow meson resonances, all associated with the presence of a heavy flavour content (charm or bottom), have attracted a lot of theoretical interest especially in view of the peculiar fact that the mass of some of them is amazingly  near to the mass of the lowest allowed mesonic decay channel~\footnote{It is impossible to mention here the enormously large number of papers on the subject. Recent excellent reviews are~\cite{Karliner:2017qhf,Liu:2019zoy}.}. 

In 2015 LHCb~\cite{Aaij:2015tga} announced the discovery of a narrow baryonic state, denoted $P_c(4450)$, with flavour content $[cd \bar c uu]$, later confirmed in~\cite{Aaij:2016ymb,Aaij:2019vzc}. However, recent photoproduction experiments ($\gamma p\to J/\psi p$) by the Gluex collaboration do not seem to confirm the existence of the $P_c(4450)$ resonance~\cite{CONF}. 

\subsection{The physics of Baryonium: a summary}

Notwithstanding the often contradictory experimental situation, quite a number of theoretical interpretations have been proposed along the years aiming at understanding in QCD the dynamics and the spectrum of multi-quark states~\cite{Rossi:1977cy,Jaffe:1976ih,DeRujula:1976zlg}~\cite{Maiani:2004uc,Rossi:2004yr,Karliner:2015ina}. 

In this talk we will focus on the scenario advocated in~\cite{Rossi:1977cy} and developed in~\cite{Rossi:2016szw} on the basis of the observation that in the large 't Hooft coupling $\lambda=g_s^2N_c$ limit, planar (in the sense specified in sect.~\ref{sec:PLDU}) diagrams dominate scattering amplitudes, and duality~\cite{DHS} can be extended from $MM \!\to\! MM$ and $B\overline B\!\to\! MM$ processes~\cite{ROSNER} to encompass the case of $B\overline B\!\to\! B\overline B$ amplitudes~\cite{Rossi:1977cy}. Within this framework in the description of the space-time evolution of a baryonic state the notion of {\it junction} naturally emerges as a running point where the three ($N_c\! =\!3$) string emanating from the three quark join to form a colour singlet. Consequently it turns out that, in order to fully specify the structure of $B\overline B\!\to \!B\overline B$ amplitudes, it is not enough to give the flow of quark flavour, but it is also necessary to provide information about the running of the junctions through the diagram. The flow of junctions in $B\overline B\!\to\! B\overline B$ diagrams allows to neatly identify and characterize annihilation and scattering amplitudes which are found to be dual to each other. 

As we said, in the large-$\lambda$ limit, duality is an exact property of QCD since planar diagrams (see sect.~\ref{sec:PLDU}) dominate. Working out the set of constraints emerging from duality, one finds that besides ordinary $q \bar q$ mesons - states with no junction - and $qqq$ ($\bar q\bar q\bar q$) baryons (anti-baryons) - states with one junction (anti-junction) - necessarily $M_4^J\!=\![qq \bar q \bar q]$ tetra-quarks - states with one junction and one anti-junction - as well as $P_5^J\!=\![qq \bar q qq]$ ($\bar P_5^J\!=\![\bar q\bar q q \bar q \bar q]$) penta-quarks (anti-penta-quarks) - states with two junction and one anti-junction (two anti-junction and one junction) - plus possibly other states with more complicated quark structure, must exist.

\begin{figure}[htb]
\vspace{9pt}
\includegraphics[scale=0.33]{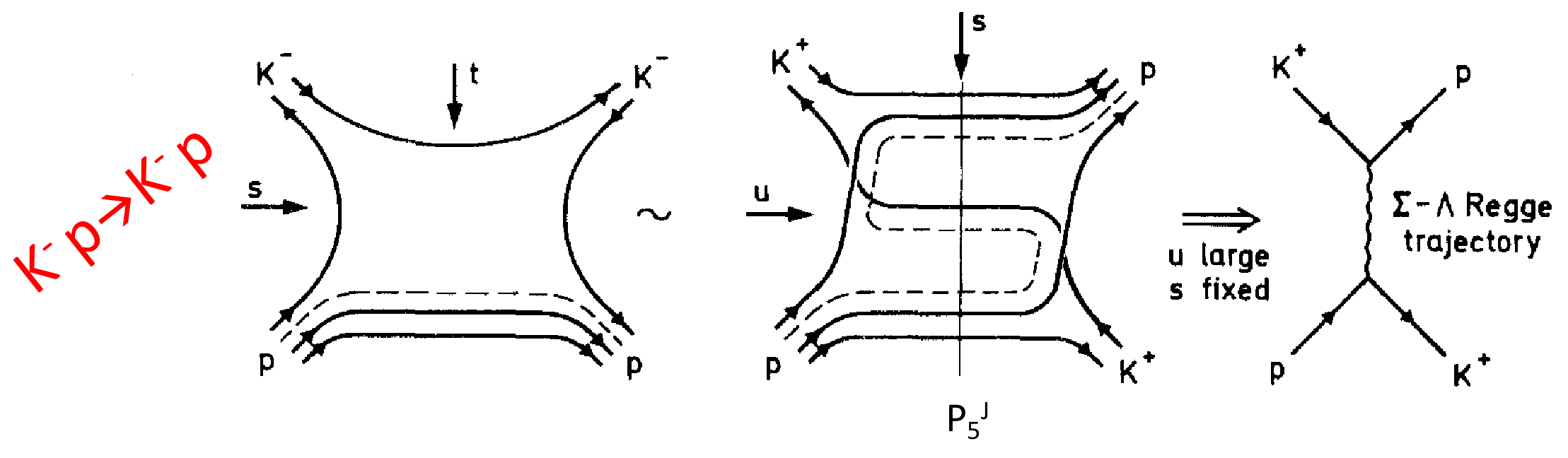}
\caption{In $K^- p\to K^- p$ penta-quarks appear as intermediate $u$-channel states dual to $B$ in $s$-channel and $M$ in $t$-channel.}
\label{fig:RTSTU}
\end{figure}

For instance, $s$-channel tetra-quarks, coupled to $B\overline B$, are dual to one-meson exchange in the $t$-channel (top-left panel of fig.~\ref{fig:fig3}). Another interesting example of duality is illustrated in fig.~\ref{fig:RTSTU}. Penta-quark states need to appear in the $u$-channel in $K^- p\!\to\! K^- p$ dual to $B\!=\!qqq$ in $s$-channel and $M\!=\!\bar q q$ in $t$-channel in order to explain breaking of exchange degeneracy in strange Regge trajectories. 

At leading order in $\lambda$ all the hadrons are stable, but allowing for quark loops they can decay by what in QCD is called a color-string breaking process. Thus the dynamically favourite decays of tetra-quarks and penta-quarks are in baryonic channels, namely  $M_4^J\to B\overline B$ and $P_5^J\!\to\! B\overline B B$, respectively. If phase space does not allow for a fully baryonic decay, the less favoured $M_4^J\!\to \!MM$ and $P_5^J\!\to\! B M$ transitions will occur with junction--anti-junction annihilation and color rearrangement. According to the baryonium scenario, lack of phase space is the reason why the width of many multi-quark states is unusually small on the $\Lambda_{QCD}$ scale. An indirect  confirmation of this interpretation comes from the discovery of the large $Y(4630)$ resonance with $\Gamma\! \sim \!92$~MeV~\cite{Pakhlova:2008vn} seen to decay in the fully baryonic $\Lambda_c^+\Lambda_c^-$ channel.

In the following we explain how all of this is born out starting from the planar, large-$\lambda$ limit of QCD, and we discuss a couple of phenomenological applications. Naturally the key issue is whether or to what extent the nice properties of planar QCD remain true in actual QCD, where $\lambda$ runs from O(1) in soft hadron physics to vanishingly small values in deep UV processes.

\section{Planarity and duality}
\label{sec:PLDU}

Extending arguments first developed in~\cite{Rossi:1977cy}, it was proved in~\cite{Rossi:2016szw} that the large strong coupling expansion of QCD is actually a large-$\lambda$ expansion. In this limit meson and baryon propagators and their scattering amplitudes are dominated by planar (in the sense specified below) diagrams. Without entering into the technicalities of the proof, we illustrate the situation with the help of a few figures~\cite{Rossi:2019szt}.

\subsection{Propagators}
\label{sec:PROP}

\begin{figure}[htb]
\begin{center}
\hspace{-.5cm}{\includegraphics[height=0.7\linewidth]{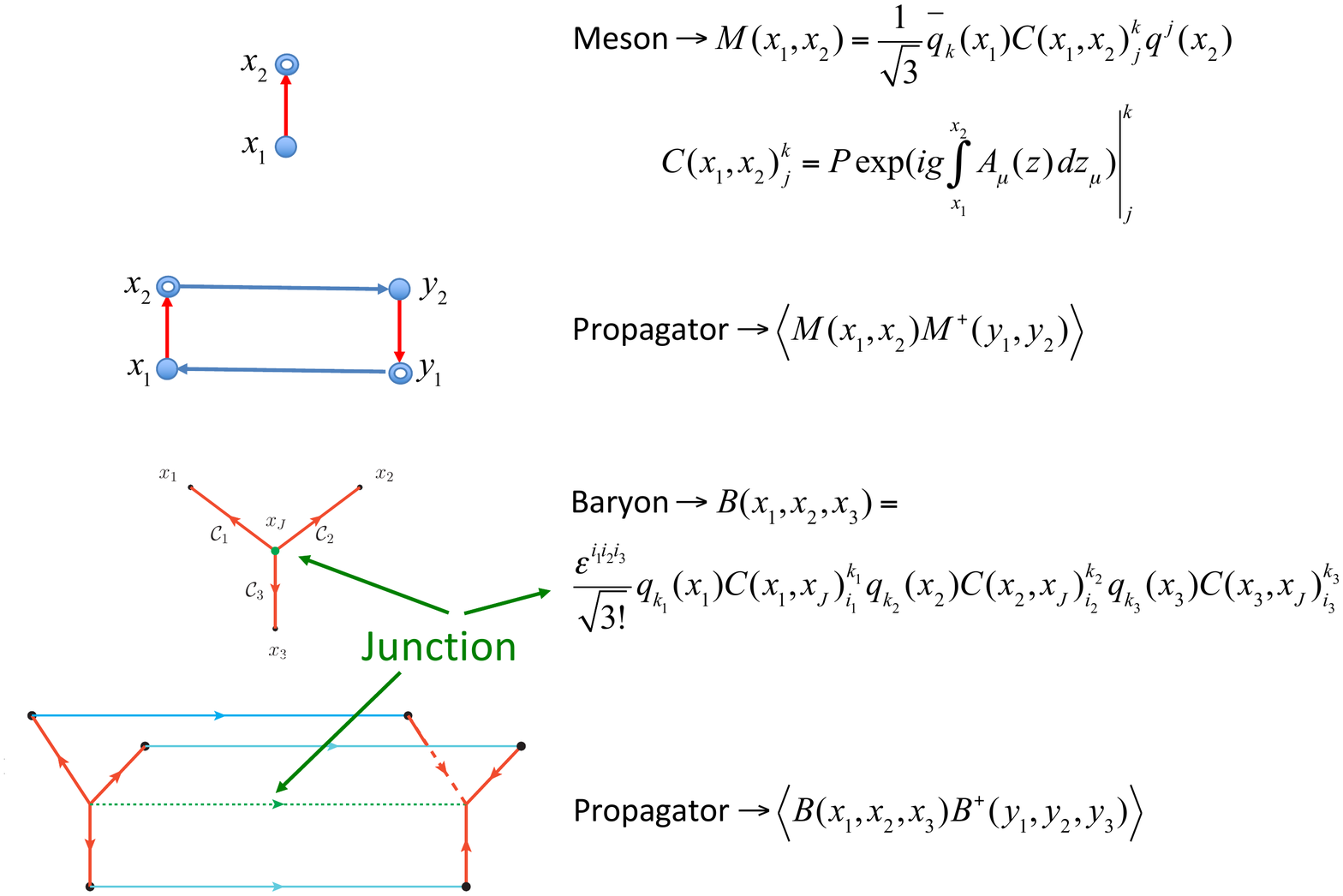}}
\caption{\it Meson and baryon propagators in the large-$\lambda$ limit.}
\label{fig:fig1}
\end{center}
\end{figure}

We display in fig.~\ref{fig:fig1} the gauge invariant expression of meson (two upper panels) and baryon (two lower panels) states and the large-$\lambda$ geometry of the world-sheet spanned by the propagators. The blue lines represent the propagation of the quarks while the red segments are color flux lines. As is well known, in this limit meson and baryons are described in terms of color strings gauge invariantly attached to quarks and/or anti-quarks.

A crucial consequence of the emerging topology of the diagrams describing the baryon propagation is the dynamical appearance of a special point (called junction~\cite{Rossi:1977cy}) where the three ($N_c=3$) Wilson lines departing from the three quarks join to make up a gauge invariant operator. Thus in the large-$\lambda$ approximation one can identify along the baryon propagator a line (drawn in green) that describes the space-time color flow, or more physically the baryon number flow. 

This suggestive picture is beautifully confirmed by some of the existing lattice studies~\cite{Bissey:2006bz,Koma:2017hcm}. The simulations of ref.~\cite{Bissey:2006bz} 
show that for sufficiently elongated strings color flux tubes get formed that connect the three quarks in the $Y$-shaped arrangement depicted in fig.~\ref{fig:fig1}.

Further evidence for the $Y$-shaped picture of baryons comes from the investigation reported in~\cite{Koma:2017hcm}. These authors are able to numerically determine the salient features of the three-quark potential showing that the latter is such that along the trajectory the propagating junction lies at a point minimizing the sum of the distances from the three quarks, i.e.\ at the solution of the three-point Fermat--Torricelli problem~\cite{FT}. This is precisely what is predicted to occur in the ``planar'' $N_c$-pages-bound-book structure describing the baryon propagation shown in fig.~\ref{fig:fig1}.

\subsection{Amplitudes}
\label{sec:AMPL}

\begin{figure}[htb]
    \begin{center}
        {\includegraphics[height=0.35\linewidth]{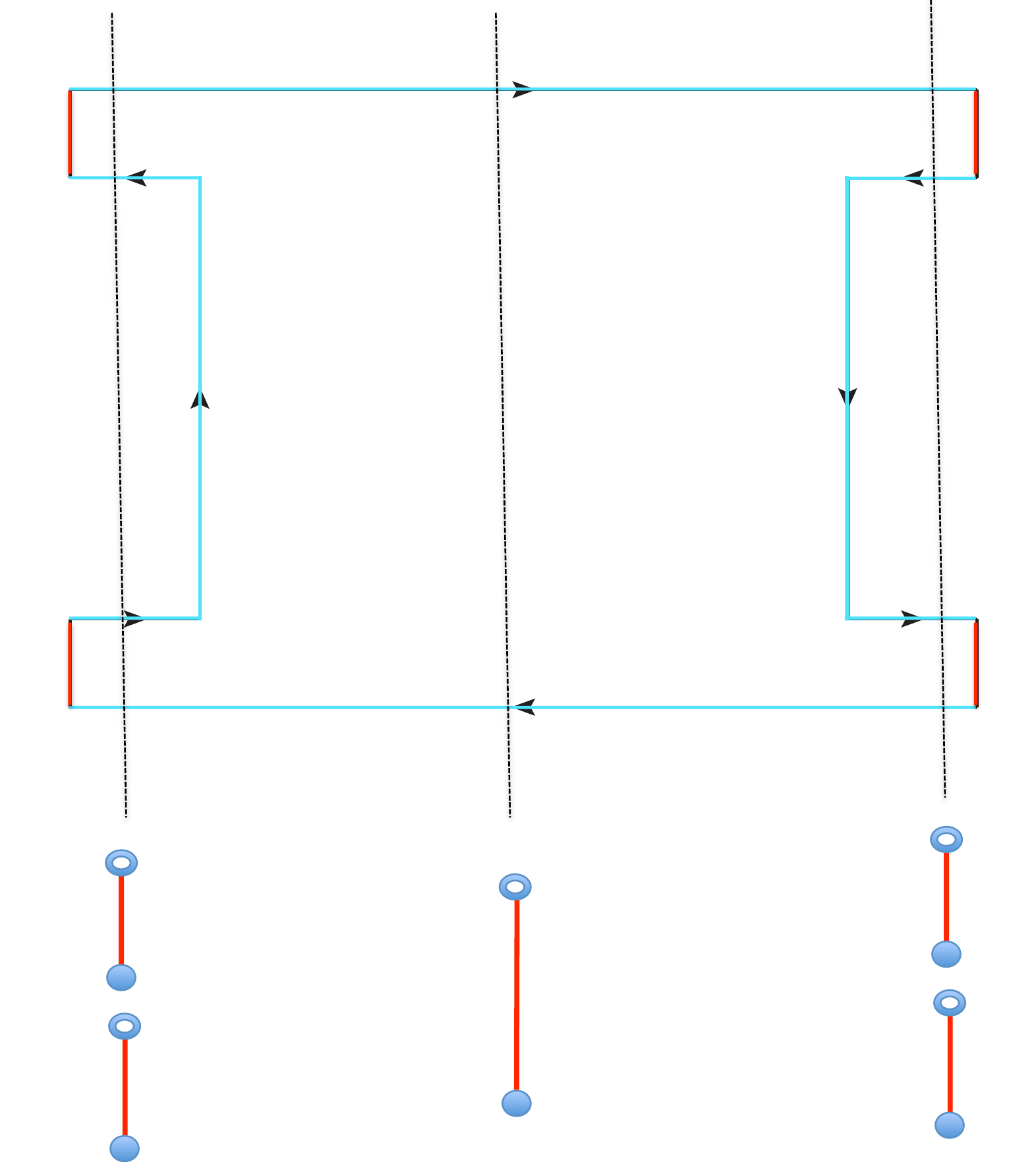}}\hspace{.3cm}
        {\includegraphics[height=0.35\linewidth]{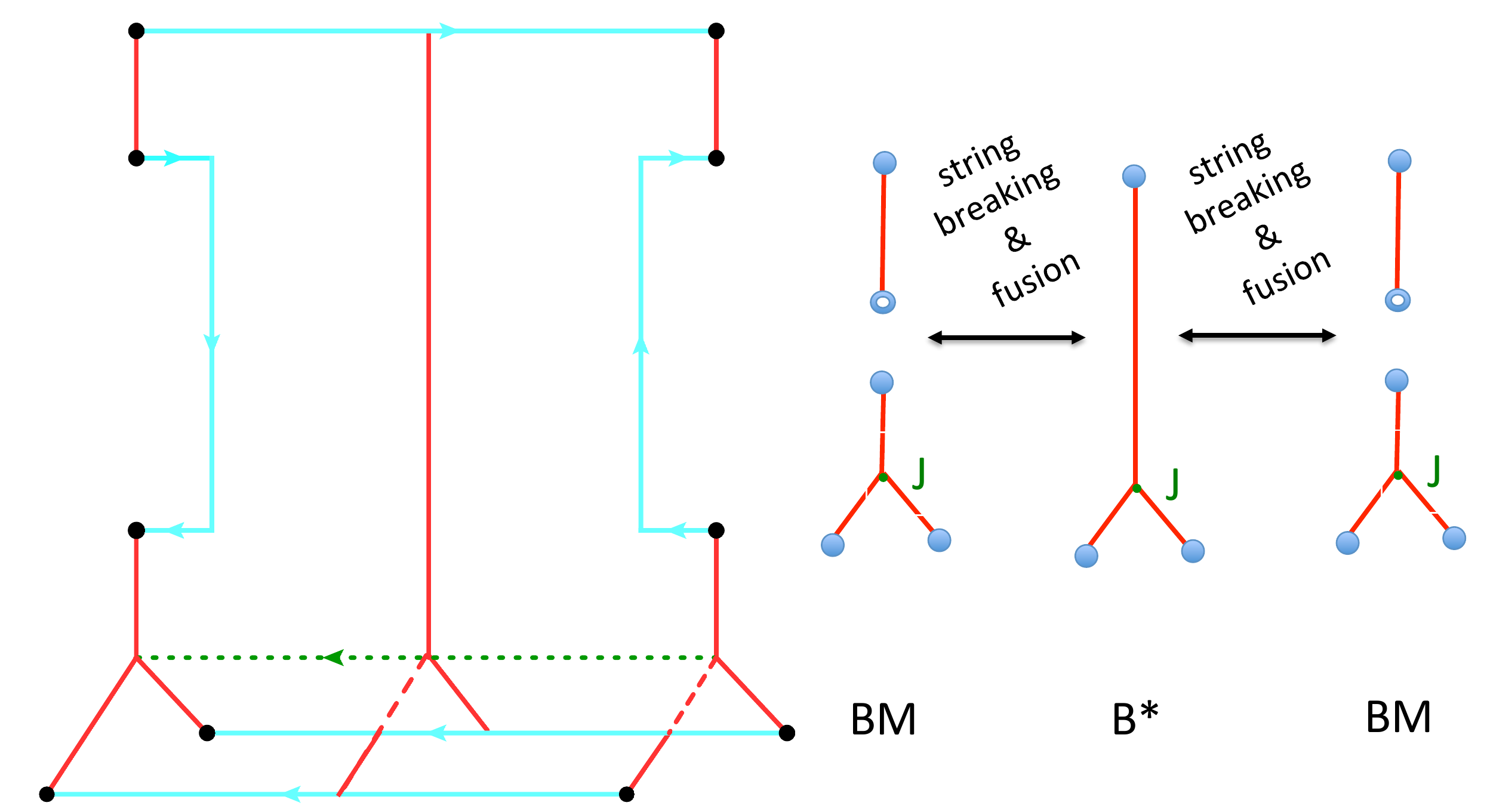}}
        \caption{\it Left: $MM\to MM$ scattering amplitude - Right: $MB\to MB$ scattering amplitude.}
\label{fig:fig2}
    \end{center}
\end{figure}

Glueing together meson strips and appropriately stretching the resulting surface, one gets the planar topology of fig.~\ref{fig:fig2} describing the $MM\!\to\! MM$ scattering amplitude to leading order in $\lambda$. The key observation is that meson-meson fusion and decay occur through color string fusion and breaking processes, respectively. 

Glueing meson to baryon sheets leads to the $MB\!\to\! MB$ amplitude depicted in the second panel of fig.~\ref{fig:fig2} where we also display the string fusion and breaking processes taking place in the $s$-channel. 

\begin{figure}[htb]
    \begin{center}
        {\includegraphics[height=0.5\linewidth]{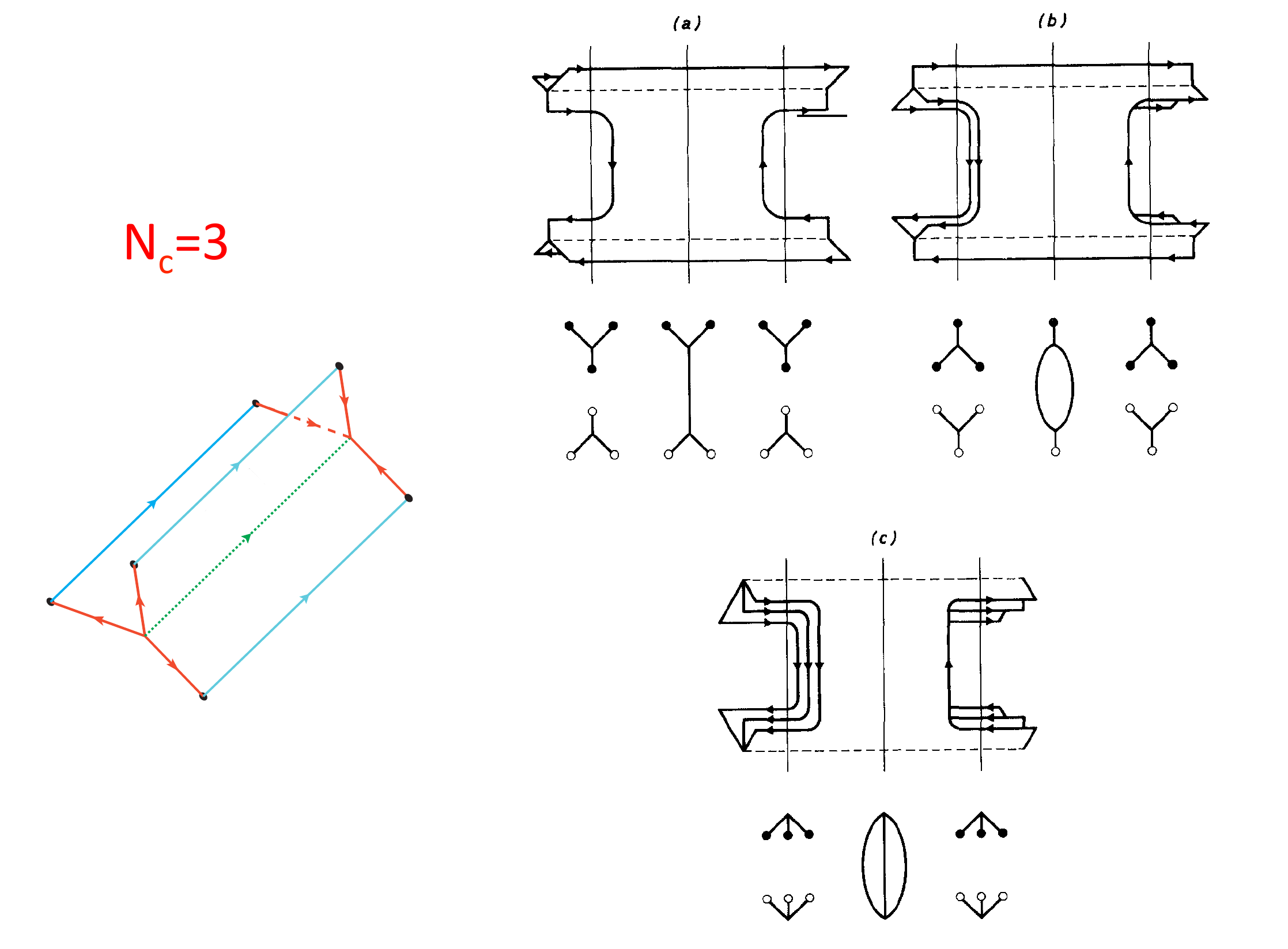}}\hspace{.5cm}
       {\includegraphics[height=0.3\linewidth]{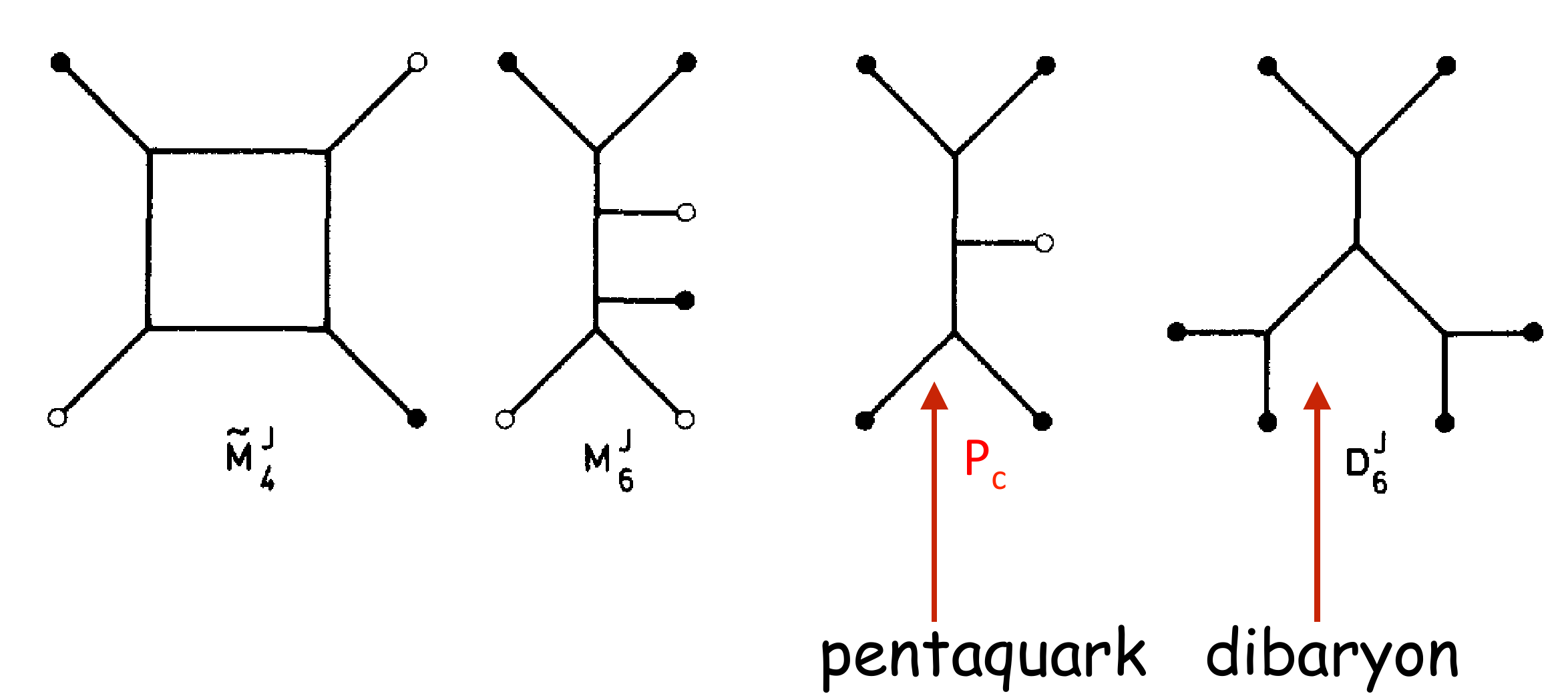}}
 \caption{\it Top: $B\bar B \to B\bar B$ scattering amplitudes - Bottom: members of the baryonium family.}
\label{fig:fig3}
    \end{center}
\end{figure}

Finally $B\bar B\!\to\! B\bar B$ amplitudes are constructed by glueing the sheets of the two $N_c$-pages-bound-books representing the propagating baryons. In the actual physical case where $N_c\!=\!3$ we have three ways of doing so, displayed in the upper part of fig.~\ref{fig:fig3}. New $s$-channel states endowed with a junction ($J$) and an anti-junction ($\overline J$) are formed. Among them we find states made by two quarks and two anti-quarks (tetra-quarks) that we have denoted by $M_4^J$. 

Besides the $B\bar B\!\!\to\! \!B\bar B$ {\it scattering} amplitudes displayed in the figure one should add three more kinds of diagrams describing {\it annihilation}. They are simply obtained from the previous ones by a $90^\circ$ rotation. The difference between the two sets of diagrams is that, while in the former the $J$ and $\overline J$ lines ($B$ and $\overline B$ baryon number) flow from the initial to the final state, in the latter the $J$ and $\overline J$ lines annihilate in the initial state, giving raise to (jets of) ordinary mesons as intermediate $s$-channel states, from which a $J\,\overline J$ pair (i.e.\ a $B\overline B$ pair) is finally created.  Following the flux of flavour is thus not enough to fully identity the structure of a diagram: one also has to specify the fate of the junctions.

From more complicated amplitudes more complicated hadrons endowed with junctions and/or anti-junctions emerge as possible intermediate states. They can be easily constructed from gauge invariance. A few of them are shown in the lower panel of fig.~\ref{fig:fig3}. Members of the baryonium family, like dibaryons and penta-quarks, were firstly conjectured to exist in~\cite{Montanet:1980te}~\footnote{A whole family of multi-quark states is also predicted in the holography inspired stringy hadron model~\cite{Sonnenschein:2016ibx}.}.

\section{A bit of phenomenology}

Though the scheme outlined in sect.~\ref{sec:PLDU} is based almost exclusively on general  geometrical considerations (symmetries \& topology), it provides a theoretical framework whereupon one can derive some useful phenomenological consequences~\cite{Rossi:1977dp,Rossi:2004yr}. As examples in the next two subsections we discuss the possibility of establishing a mass formula to estimate the mass of certain narrow multi-quark states and we illustrate the large-$\lambda$ description of $P_c$ photoproduction that is an interesting process for hidden flavour penta-quark production.

\subsection{A mass formula}

With an eye to the supposedly dynamically dominant string breaking decay mode of baryonium, we may be tempted to test the oversimplified mass formulae
\begin{eqnarray}
&& M^{J\,{\overline{ \! J}}}_{qq\bar q \bar q} +\Delta_{HSB} =M_{B_1}+M_{\overline{B_2}} \, ,\label{MPM}\\
&& M^{J\,{\overline{ \! J}}J}_{qq\bar q qq} + 2\Delta_{HSB} =M_{B_1}+M_{\overline{B_2}} +M_{B_3} \, ,\label{MP}
\end{eqnarray}
with $\Delta_{HSB}$ the ``hadronic string breaking'' contribution, against existing narrow tetra- and penta-quarks with mass below the corresponding $B\overline B$ or $B\overline B B$ threshold~\footnote{More sophisticated multi-quark mass formulae also based on the general notion of junction have been successfully proposed in~\cite{Karliner:2014gca,Karliner:2020vsi}.}.

We (arbitrarily) take $\Delta_{HSB}$ from the mass of the $X(6900)$ tetra-quark with $[cc \bar c \bar c]$ flavour content~\cite{Aaij:2020fnh} finding a value well in line with QCD expectations, i.e.~\footnote{All masses are expressed in MeV.}
\begin{equation}
\hspace{-.6cm}\Delta_{HSB}=2M_{ccu}-6900=7242-6900 \sim 350\, .\label{DEL}
\end{equation}

\subsubsection{A few applications}

With this information we can use eqs.~(\ref{MPM}) and~(\ref{MP}) to give an order of magnitude estimate of the mass of a few (conjectured) tetra- and penta-quark (narrow) states.

1) For the mass of the doubly charmed $[cc\bar u\bar u]$ tetra-quark, not yet identified, we get the ``pre-diction''
\begin{equation}
\hspace{-.8cm}M^{J\,{\overline{ \! J}}}_{cc\bar u \bar u}\! \! =\! M_{ccu}\! \!+\! M_{\bar u\bar u\bar u} \! -\! \Delta_{HSB}\! =\!  3621\! +\! 1236\! -\! 350\! \sim \! 4510 \label{MA1}
\end{equation}

2) For the $X_0(2900)$ tetra-quark with flavour content $[du\bar c \bar s]$~\cite{Aaij:2020hon} we get the reasonably good ``post-diction'' 
\begin{equation}
\hspace{-.7cm}M^{J\,{\overline{ \! J}}}_{ud \bar c \bar s} =M_{\bar c \bar s \bar u}+M_{u u d} -\Delta_{HSB}= 3400-350\sim 3050\label{MA2}
\end{equation}

3) For the $Z(4430)^-$ tetra-quark with flavour content $[cd\bar c \bar u]$~\cite{Aaij:2014jqa} we get the rather good ``post-diction'' 
\begin{equation}
\hspace{-.7cm}M^{J\,{\overline{ \! J}}}_{cd\bar c \bar u} = M_{cdd}+M_{\bar d\bar c\bar u}-\Delta_{HSB} = 4908-350 \sim 4558\label{MA3}
\end{equation}

4) For the $Y_b(10888)^*$ tetra-quark with flavour content $[b q\bar b \bar q]$~\cite{YB} we get the very good ``post-diction'' 
\begin{equation}
\hspace{-.7cm}M^{J\,{\overline{ \! J}}}_{b q\bar b \bar q} =2M_{bud}-\Delta_{HSB} = 11250- 350 \sim 10900 \label{MA4}
\end{equation}

5) The $P_c(4450)$~\cite{Aaij:2015tga} mass estimate provided by eq.~(\ref{MP}) is not particularly good. One gets in fact
\begin{equation}
\hspace{-.7cm}P^{J\,{\overline{\! J}} J}_{uc \bar c ud} \!=\!2M_{ucd}\!+\!M_{u u d} \!-\!2\Delta_{HSB}\!=\! 5510\!-\!700\!\sim \!4810 \label{PC2}
\end{equation}

\subsection{$P_c$ photoproduction}

$P_c[uc\bar c ud]$ is a hidden charm state. So it is natural to think at photoproduction on a proton target as an efficient way to create a $c\bar c$ pair (or for that matter any other flavour pair).

\begin{figure}[htb]
\begin{center}
{\includegraphics[height=0.4\linewidth]{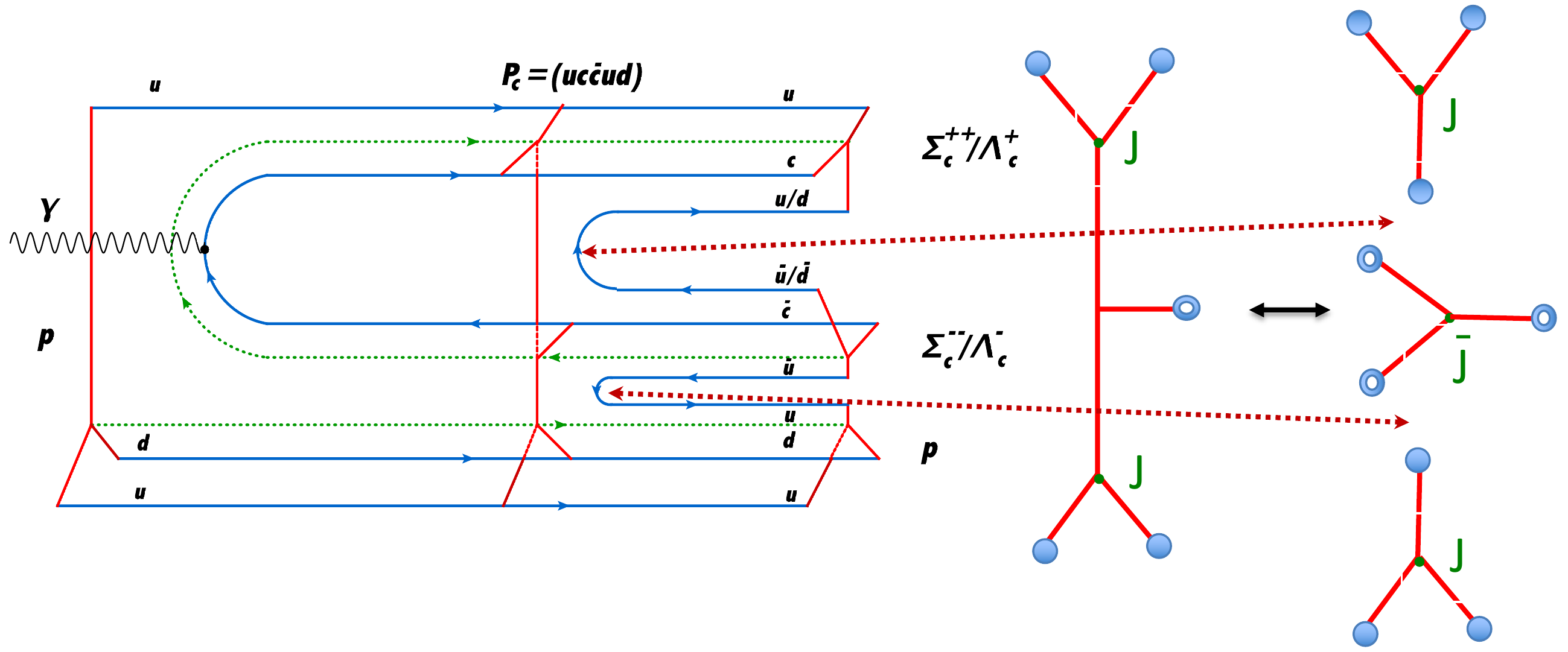}}
\caption{\it The kinematically forbidden $P_c\to B\bar B B$ decay. The arrows indicate the string breaking points.}
\label{fig:fig5}
\end{center}
\end{figure}

Were the $P_c$ mass larger than the three-baryon threshold~\footnote{The fully baryonic decays $P_c\!\to\! \Lambda_c^{+}\!+\!\Lambda_c^{-}\!+\!p$ and $P_c\!\to\! \Sigma_c^{++}\!+\!\Sigma_c^{--}\!+\!p$ are kinematically forbidden, because $M_{P_c}\!\sim 4450$ while $2 M_{\Lambda_c^{+}}\!+\!M_p\!\sim 2 \times 2286 + \!938\!=\! 5510$ and $2 M_{\Sigma_c^{++}}\!+\!M_p\!\sim\!2\times\! 2454 \!+ \!938\!= \!5846$.} the typical dynamically dominant (i.e.\ ``planar'') diagram would be the one shown in fig.~\ref{fig:fig5} where the intermediate $P_c$ resonance decay proceeds via two string breakings.

\begin{figure}[htb]
    \begin{center}
        {\includegraphics[height=0.42\linewidth]{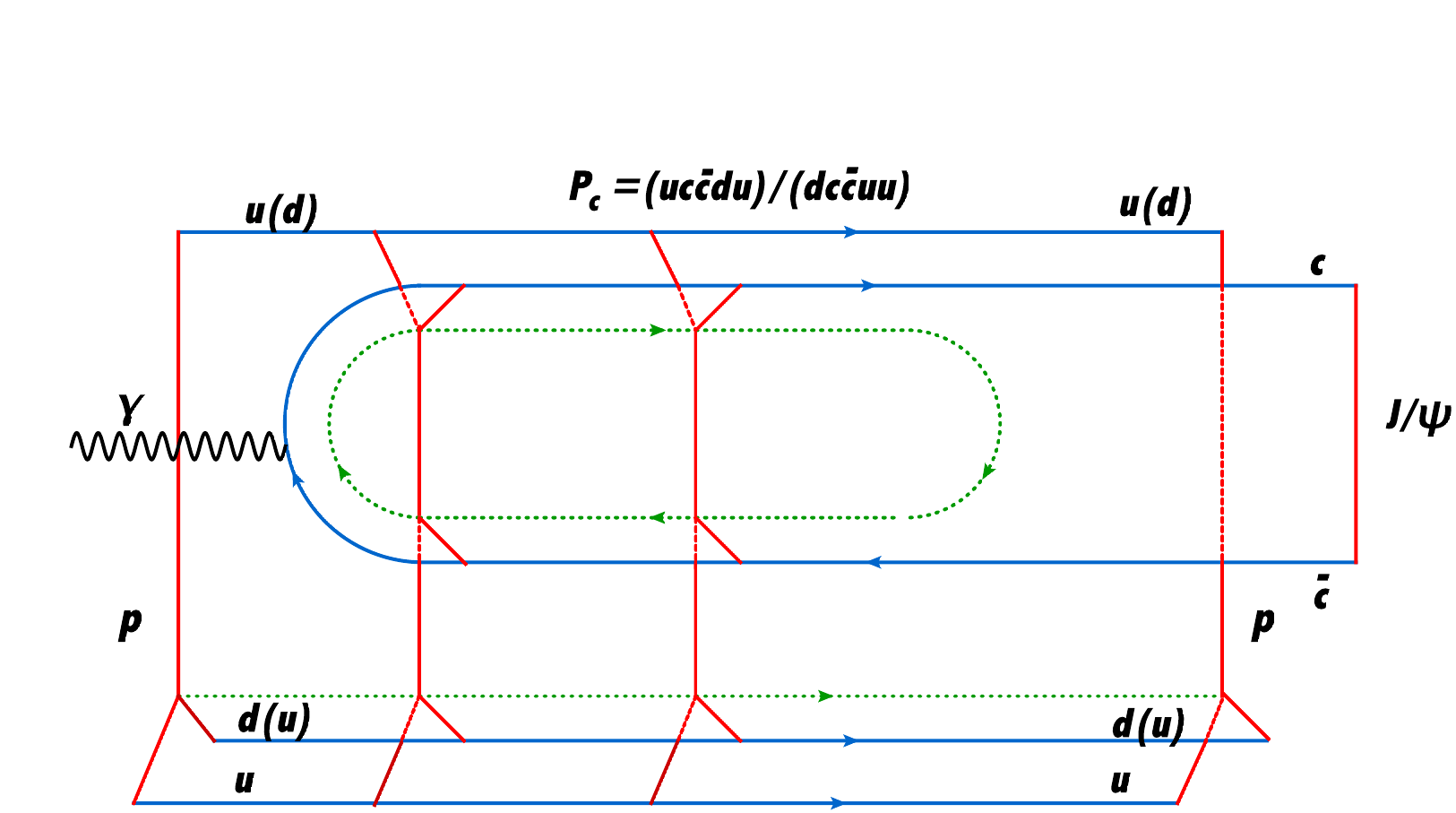}}\vspace{.5cm}
       {\includegraphics[height=0.34\linewidth]{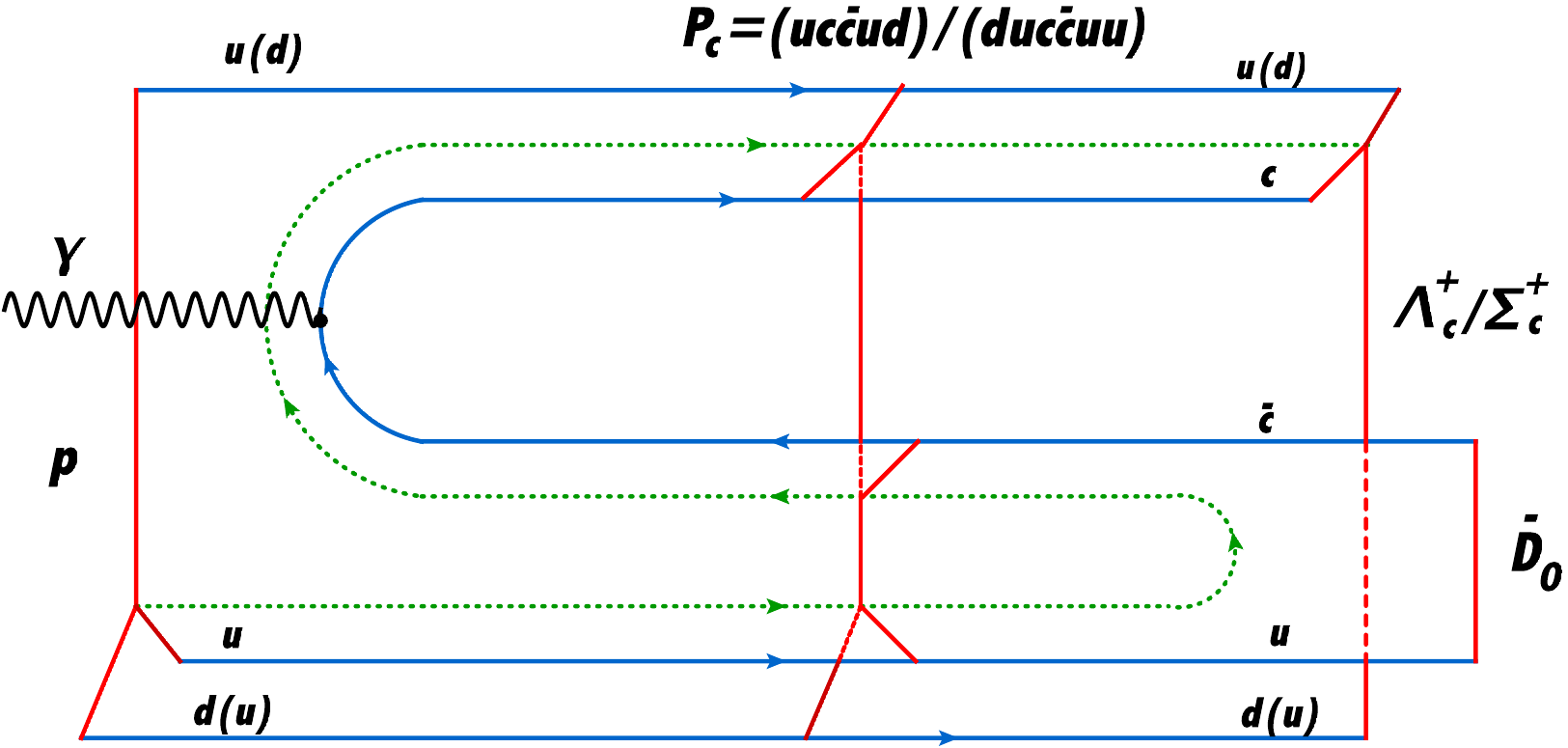}}\vspace{.5cm}
       {\includegraphics[height=0.34\linewidth]{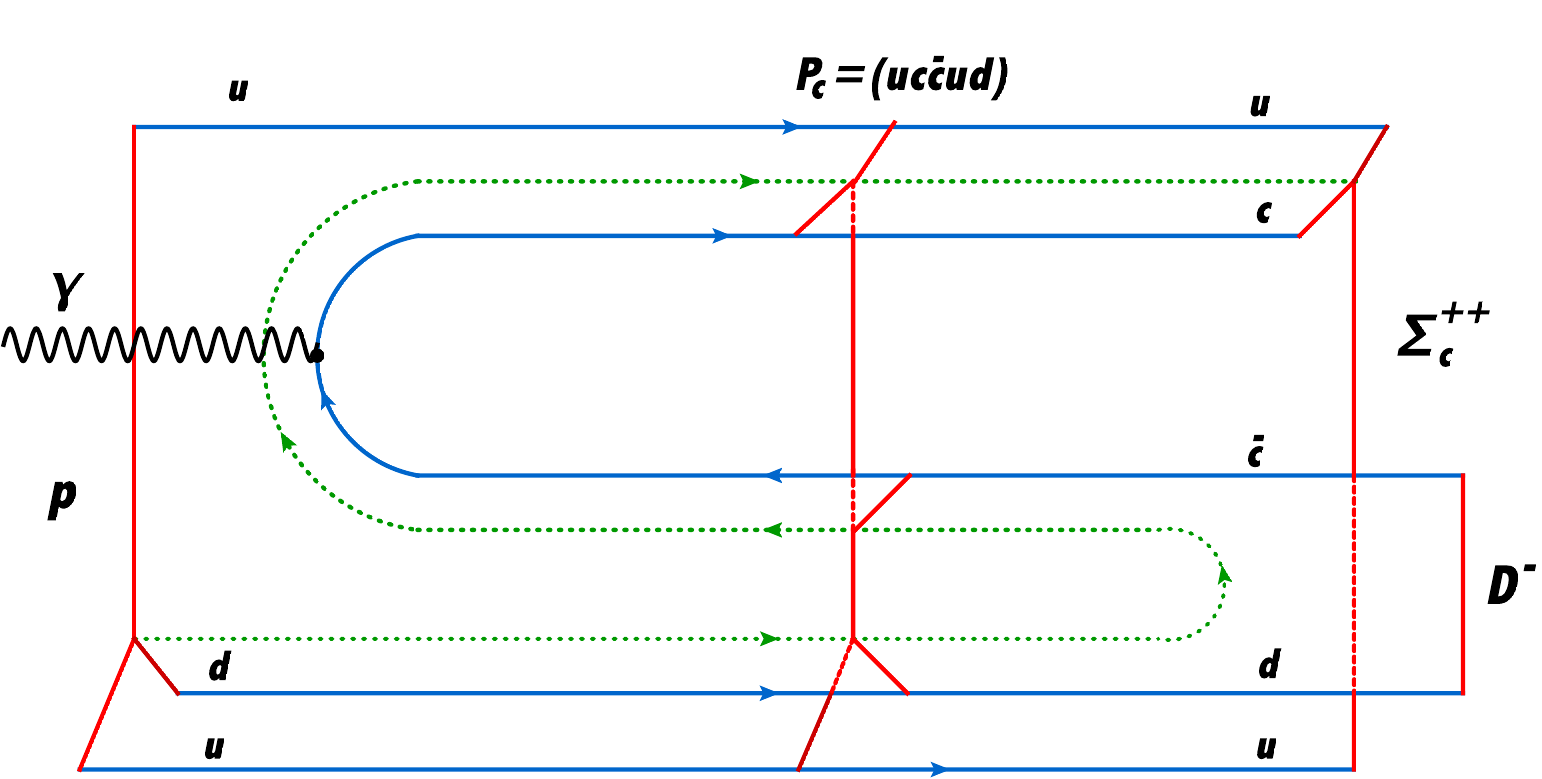}}
 \caption{\it Colour rearrangement $P_c$ decay. Top: $P_c\to J/\psi + p$ - Middle: $P_c\to \Lambda_c^+ /\Sigma^+_c + \bar D_0$ - Bottom: $P_c\to \Sigma^{++}_c + D^-$.}
\label{fig:fig6}
    \end{center}
\end{figure}

This process is, however, kinematically forbidden and the decay needs to proceed via color rearrangement, i.e.\ via a suppressed $J-\bar J$ annihilation. Examples of such decays are $P_c\!\to\! J/\psi \!+\! p$, $P_c\!\to\! \Lambda_c^+ /\Sigma^+_c \!+\! \bar D_0$ and $P_c\!\to\! \Sigma^{++}_c \!+\! D^-$. They are depicted in fig.~\ref{fig:fig6}. At the moment we have no way to establish the relative magnitude of the various decay branching ratios. We can only notice the different topology of the top diagram with respect of the other two. In the top diagram two junctions are closed in a loop (``bathtub'' diagrams), while in the other two the junction line unrolls through the diagram (``snake' diagrams). In the Appendix we explain how different colour contractions give raise to these two different topologies.

\vspace{-.2cm}
\subsubsection{Tagged photoproduction}
\label{sec:TAPH}

\begin{figure}[htb]
    \begin{center}
        {\includegraphics[height=0.4\linewidth]{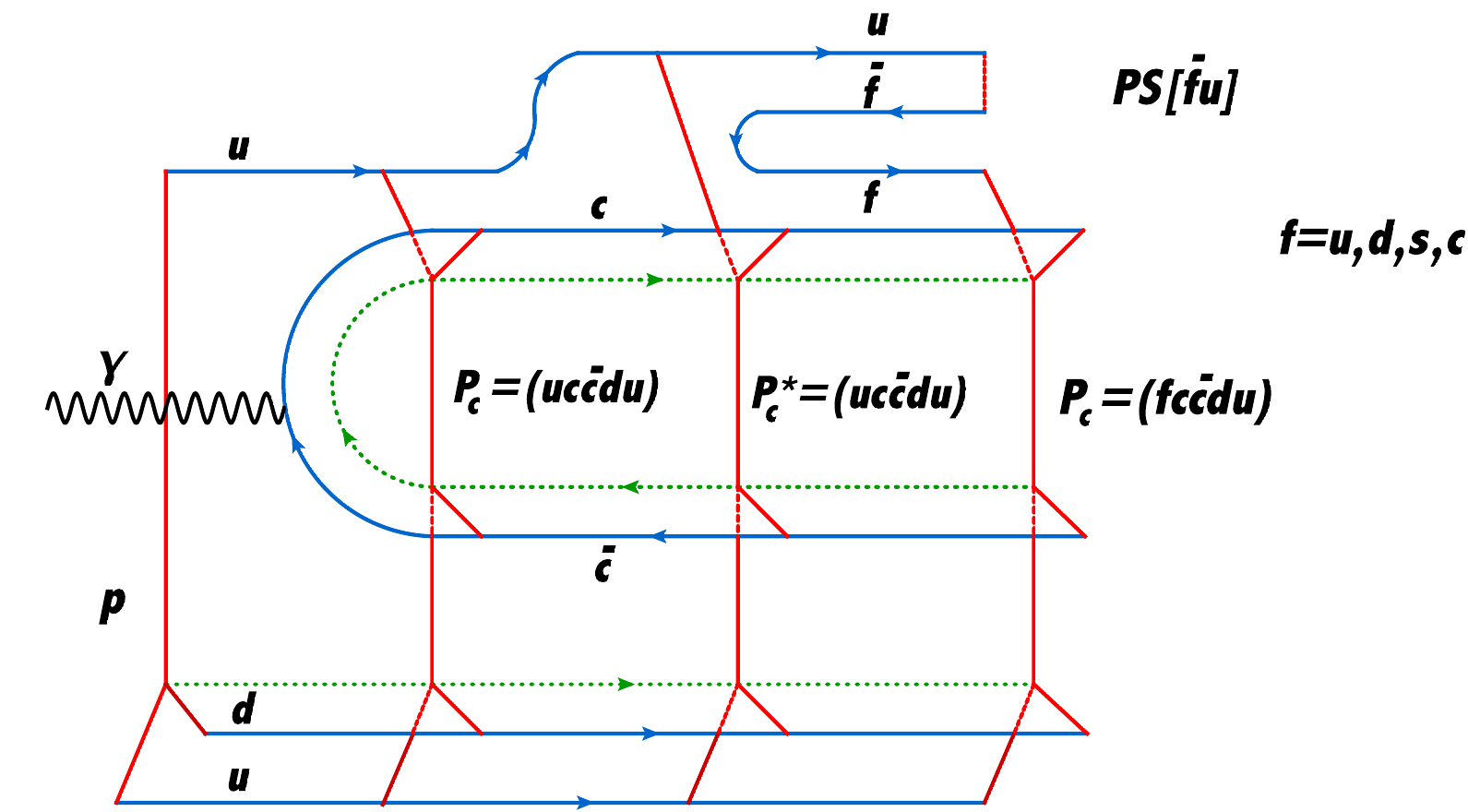}}\vspace{.5cm}
       {\includegraphics[height=0.4\linewidth]{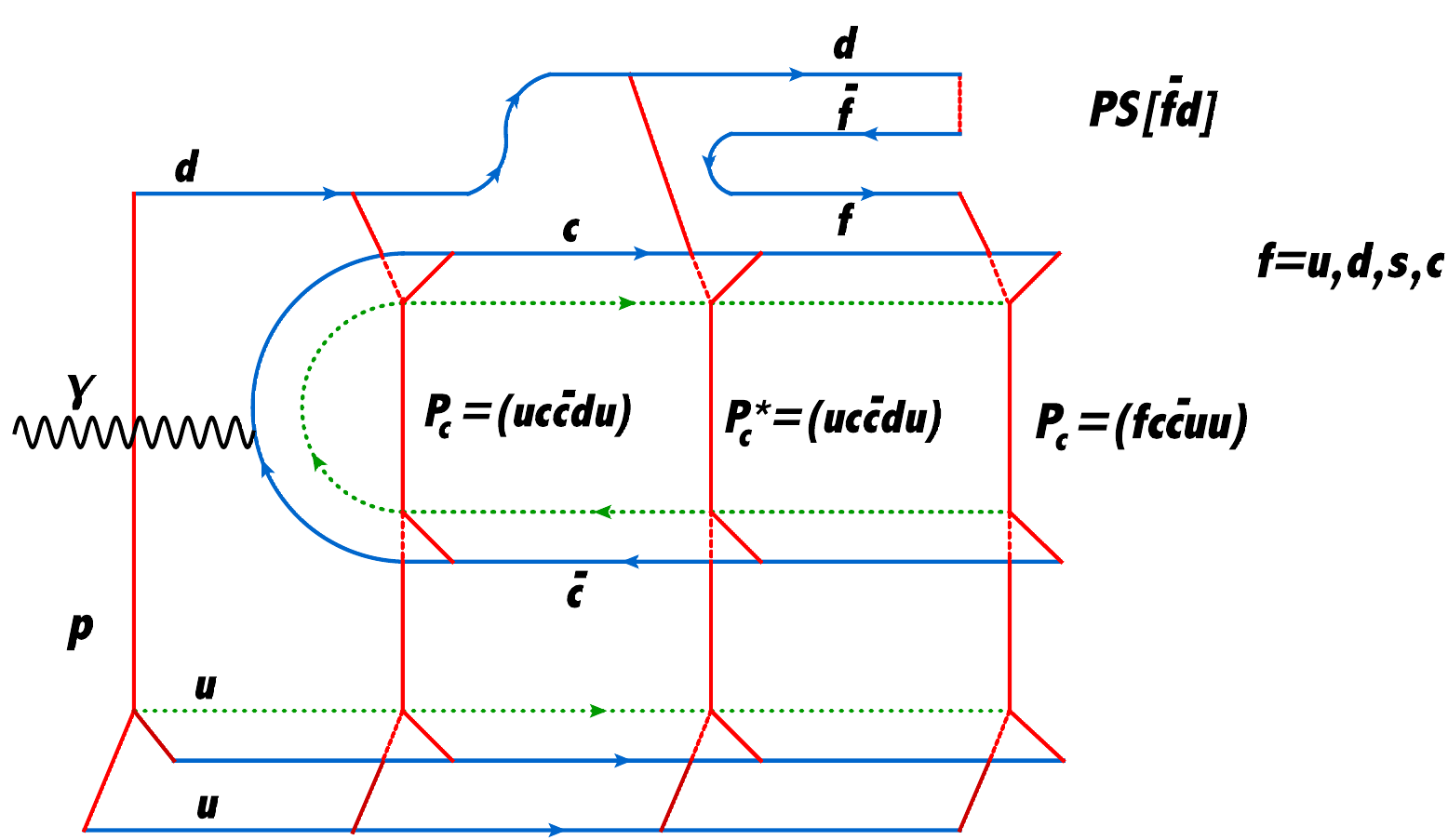}}
 \caption{\it Top: $\gamma p\to P_c^\star [f c \bar cud]\to  PS[\bar f u] + P_c[f c \bar c ud]$ - Bottom: $\gamma p\to P_c^\star [f c \bar c u u]\to PS[\bar f d] +\,P_c[f c \bar c uu]$.}
\label{fig:fig7}
    \end{center}
\end{figure}

If sufficiently energetic photons are available, the pattern $P[fc \bar c ud]$ of the $P_c$ flavour partners can be studied with the help of a tagged production process where a PS meson $\bar f u/\bar f d=\pi, K, D$ is identified, as seen in fig.~\ref{fig:fig7}.

\subsubsection{Peak resolution}\label{sec:PRES}

The recent resolution of the $P_c$ peaks~\cite{Aaij:2019vzc} indicates the presence of three resonances that have a similar interpretation in our picture as in the diquark model of ref.~\cite{Maiani:2015vwa} (for an alternative scenario see~\cite{Karliner:2015ina}). In our scheme the lighter peak (4312~MeV) should be identified with the $(ud)_{I=0}\, \bar{c}\, (uc)$ state, while the two heavier and almost degenerate ones (4450 \& 4457~MeV) with $(ud)_{I=1} \,\bar{c} \,(uc)$ \& $(uu)_{I=1}\, \bar{c} \,(dc)$ (see fig.~\ref{fig:fig6}).

\section{Conclusions}

We have shown in this talk that planarity and the QCD notion of colour string provide a unified description of ordinary hadrons (mesons and baryons) as well as other multi-quark states (tetra-, penta-quarks, $\ldots$) and of their scattering amplitudes. 

In the limit of large $\lambda\!=\!g^2_sN_c$ ('t Hooft coupling), planarity is an exact property of QCD which in turn implies duality. In this limit the notion of junction naturally emerges as a characteristic property of baryons and of a new family of multi-quark hadrons. States endowed with an even number of junctions $\!+\!$ anti-junctions are states with vanishing baryon number. States endowed with an odd number of junctions $\!+\!$  anti-junctions are states with a non-vanishing baryon number. 

If we can assume that the dominant decay of these states is via string breaking, then the dynamically favourite decay of tetra-quarks (states hosting a $J\!-\!\bar J$ pair) is in $B \overline B$ and that of penta-quarks (states hosting a $J\!-\!\bar J \!- \!J$ triplet) in $B \overline B B$ (anti-penta-quarks, states hosting a $\bar J\!-\!J \!-\!\bar J$ triplet, is in $\overline B B \overline B$). In such processes the number of junctions and anti-junctions is separately conserved. 

The family encompassing all these states is called ``baryonium'', because its tetra-quark members can be seen as hidden baryon number states, much like with charmonium (bottomonium) we denote resonances with hidden charm (bottom). 

A striking phenomenological observation is that many baryonium states are surprisingly narrow with mass over width ratio $M/\Gamma \!= \!{\mbox O}(10^{-2})$, making them visible in experiments. In our scenario this unusually small width is a consequence of the fact that, if the dynamically favoured string-breaking decay is kinematically forbidden, only the decays in which the disfavoured junction-anti-junction annihilation processes occurs can take place.

Naturally the key question is to what extent in actual QCD the nice large-$\lambda$ properties we have summarized above remain valid. Most probably actual multi-quark states are complicated combinations of baryonium-like resonances~\cite{Rossi:1977cy}, diquark--anti-diquark bound states~\cite{Jaffe:1976ih,Maiani:2015vwa} and molecular di-meson configurations~\cite{DeRujula:1976zlg}\cite{Karliner:2015ina}. 

As a couple of more phenomenological applications of the baryonium paradigma, we have 1)  discussed a simple formula capable of providing the order of magnitude of the mass of some narrow baryonium states and 2) illustrated the virtues of photoproduction as a tool to produce hidden flavour penta-quarks, like $P_c$.

{\bf Acknowledgements -} We wish to thank the Organizers for the exciting atmosphere of  the Conference.

\appendix
\section{$\epsilon$-contractions \& color rearrangement}
\label{sec:APPA}

We collect here a few formulae relevant for $M_4^J$ and $P_5^J$ decay, involving $\epsilon$-symbol contractions.

\subsection{Two $\epsilon$-symbol contraction}
\label{sec:APPA1}

Consider for concreteness the $Q=-2$ tetra-quark 
\begin{equation}
\hspace{-.5cm}M^J_4 [\bar c \bar u-sd] = \bar c^{i_1}_w\bar u^{i_2}_w[\sum_k\epsilon_{i_1i_2k'} W_{k}^{k'}\epsilon^{k j_1 j_2}] s^w_{j_1} d^w_{j_2} \, ,
\label{B4}
\end{equation}
where $q^w_{j_1}=W_{j_1}^{j'_1}q_{j'_1}$ and $\bar q^{i_1}_w= \bar q^{i'_1}W_{i'_1}^{i_1}$ with $W_{i'}^{i}$ the Wilson line. In order to study the possible overlap of the most general flavoured tetra-quark with the two-meson channel, one needs to consider the combination
\begin{eqnarray}
\hspace{-1.cm}&&\bar q_{wf_1}^{i_1}\bar q_{wf_2}^{i_2}[\sum_k\epsilon_{i_1i_2k} \epsilon^{kj_1j_2}] q^{wg_1}_{j_1} q^{wg_2}_{j_2}=\nonumber \\
\hspace{-1.cm}&&=\bar q_{wf_1}^{i_1}\bar q_{wf_2}^{i_2}[\delta_{i_1}^{j_1}\delta_{i_2}^{j_2}-\delta_{i_1}^{j_2}\delta_{i_2}^{j_1}]q^{wg_1}_{j_1} q^{wg_2}_{j_2}=\nonumber\\
\hspace{-1.cm}&&= [\bar q_{wf_1}^{i} q^{wg_1}_{i}] \, [\bar q_{wf_2}^{j} q^{wg_2}_{j}] - [\bar q_{wf_1}^{i} q^{wg_2}_{i}] \, [\bar q_{wf_2}^{j} q^{wg_1}_{j}] \, ,\label{TWOCONTR} 
\end{eqnarray}
where we have shrunk to the identity the Wilson line between the two $\epsilon$'s. This drastic colour rearrangement is necessary to end up with colour singlet $q \bar q$ mesons. The result above implies for instance
\begin{equation}
M^J_4 [\bar c \bar u-sd] = [D_s^- \pi^-] - [D^- K^-] \, ,
\label{DEC}
\end{equation}
which means that in the baryonium picture the decays 
\begin{eqnarray}
&&M^J_4 [\bar c \bar u-sd] \to D_s^- +\pi^- \label{DEC1}\\
&&M^J_4 [\bar c \bar u-sd] \to D^- + K^- \label{DEC2}
\end{eqnarray}
are expected to have equal amplitude, up to phase-space effects that would tend to favour the decay~(\ref{DEC1}), as $m_{D_s^-}+m_{\pi^-}\sim 2107~{\mbox{MeV}} < m_{D^-}+m_{K^-}\sim 2364~{\mbox{MeV}}$.

\subsection{Three $\epsilon$-symbol contraction}
\label{sec:APPA2}

Consider the three $\epsilon$-symbol contraction 
\begin{equation}
C^{i_1i_2j_1j_2}_{k}=\sum_{m,n}  \epsilon^{i_1i_2m} \epsilon_{m k n} \epsilon^{nj_1j_2}
\, .\label{THREECONTR}
\end{equation}
According to whether one first performs the contraction of the first two $\epsilon$'s or the second two, one gets  
\begin{eqnarray}
\hspace{-1.3cm}&&C^{i_1i_2j_1j_2}_{k}\Big{|}^{(1)}=\sum_{m,n}  \epsilon^{i_1i_2m} \epsilon_{m k n} \epsilon^{nj_1j_2} =\nonumber\\
\hspace{-1.3cm}&& = \sum_{n}  [\delta^{i_1}_{k}\delta^{i_2}_{n}-\delta^{i_1}_{n}\delta^{i_2}_{k}]\epsilon^{nj_1j_2} 
=\delta^{i_1}_{k} \epsilon^{i_2j_1j_2} -\delta^{i_2}_{k}\epsilon^{i_1j_1j_2}  \, ,
\label{THREECONTR1}\\
\hspace{-1.3cm}&&C^{i_1i_2j_1j_2}_{k}\Big{|}^{(2)}=\sum_{m,n}  \epsilon^{i_1i_2m} \epsilon_{m kn} \epsilon^{nj_1j_2}=\nonumber\\
\hspace{-1.3cm}&&= \sum_{m} \epsilon^{mi_1i_2} [\delta_{m}^{j_1}\delta_{k}^{j_2}\!-\!\delta_{k}^{j_1}\delta_{m}^{j_2}] = \epsilon^{j_1i_1i_2}\delta_{k}^{j_2} - \epsilon^{j_2i_1i_2} \delta_{k}^{j_1} \, .
\label{THREECONTR2}
\end{eqnarray}
We thus anticipate that in the final state different flavour combinations of baryon-meson states are possible. To see this more clearly let's saturate eqs.~(\ref{THREECONTR1}) and~(\ref{THREECONTR2}) with the two flavour structures 
$d_{i_1}^wc_{i_2}^w u_{j_1}^w u_{j_2}^w\bar c_w^{k}$ and $u^w_{i_1}c^w_{i_2}u^w_{j_1}d^w_{j_2}\bar c_w^{k}$. Again, shrinking to the identity the two Wilson lines connecting the three $\epsilon$ symbols (see the bottom panel of fig.~\ref{fig:fig3}) in order to get colour singlet $q\bar q$ and $qqq$ states, we obtain the four combinations
\begin{itemize} 
\item $P_c^{(1)}= \sum_{m,n}  \epsilon^{i_1i_2m} \epsilon_{m k n} \epsilon^{nj_1j_2}d^w_{i_1}c_{i_2}^wu_{j_1}^wu_{j_2}^w\bar c_w^{k}$ 
\begin{enumerate}
\item $d^w_{i_1}c^w_{i_2}u^w_{j_1}u^w_{j_2}\bar c_w^{k} C^{i_1i_2j_1j_2}_{k}\Big{|}^{(1)}=[\bar c d][cuu] -[\bar c c][duu] = D^-\Sigma_c^{++}-J/\psi \, p$
\item $d^w_{i_1}c^w_{i_2}u^w_{j_1}u^{'w}_{j_2}\bar c_w^{k} C^{i_1i_2j_1j_2}_{k}\Big{|}^{(2)}= [\bar c u][cdu'] - [\bar c u'][cdu] = \bar D^0\,\Lambda_c^+/\Sigma_c^+-\bar D^0\,\Lambda_c^+/\Sigma_c^+$
\end{enumerate}
\item $P_c^{(2)}= \sum_{m,n}  \epsilon^{i_1i_2m} \epsilon_{m k n} \epsilon^{nj_1j_2}
u^w_{i_1}c^w_{i_2}u^w_{j_1}d^w_{j_2}\bar c_w^{k} $ 
\begin{enumerate}
\item $u^w_{i_1}c^w_{i_2}u^w_{j_1}d^w_{j_2}\bar c_w^{k} C^{i_1i_2j_1j_2}_{k}\Big{|}^{(1)}=[\bar c u][cdu]-  [\bar c c][duu] = \bar D^0\,\Lambda_c^+/\Sigma_c^+- J/\psi \, p$
\item $u^w_{i_1}c^w_{i_2}u^w_{j_1}d^w_{j_2}\bar c_w^{k} C^{i_1i_2j_1j_2}_{k}\Big{|}^{(2)}=[\bar c u][cdu]-  [\bar c d][cuu] = \bar D^0\,\Lambda_c^+/\Sigma_c^+-D^-\Sigma_c^{++}$
\end{enumerate}
\end{itemize}
These formulae give us the possible non-fully baryonic $P_c$ decay channels. We see from them that the ``bathtub'' (top panel of fig.~\ref{fig:fig6}) is formed when the junction, which is photoproduced together with a $\bar c c$ pair, annihilates leaving a $J/\psi$ meson in the final state. This happens only in the second term of the contraction $C\,|^{(1)}$ (eq.~(\ref{THREECONTR1})) owing to the presence of the $\delta^{i_2}_{k}$ factor.

\end{document}